\DocumentMetadata{}
\documentclass[sigconf]{acmart}

\newcommand{\system}{WireBend-kit}

\usepackage{gensymb} 
\AtBeginDocument{%
  \providecommand\BibTeX{{%
    \normalfont B\kern-0.5em{\scshape i\kern-0.25em b}\kern-0.8em\TeX}}}

\copyrightyear{2025}
\acmYear{2025}
\setcopyright{rightsretained}
\acmConference[SCF '25]{ACM Symposium on Computational Fabrication}{November 20--21, 2025}{Cambridge, MA, USA}
\acmBooktitle{ACM Symposium on Computational Fabrication (SCF '25), November 20--21, 2025, Cambridge, MA, USA}
\acmDOI{10.1145/3745778.3766662}
\acmISBN{979-8-4007-2034-5/2025/11}

\begin{document}

\title[WireBend-kit]{WireBend-kit: A Computational Design and Fabrication Toolkit for Wirebending Custom 3D Wireframe Structures}




\author{Faraz Faruqi}
\affiliation{%
  \institution{MIT CSAIL}
  \streetaddress{MIT CSAIL}
  \city{Cambridge}
  \state{MA}
  \country{USA}}
\email{ffaruqi@mit.edu}

\author{Josha Paonaskar}
\affiliation{%
  \institution{University of Washington}
  \streetaddress{Guggenheim Hall}
  \city{Seattle}
  \state{WA}
  \country{USA}}
\email{joshap@uw.edu}

\author{Riley Schuler}
\affiliation{%
  \institution{University of Washington}
  \streetaddress{Guggenheim Hall}
  \city{Seattle}
  \state{WA}
  \country{USA}}
\email{rileys28@uw.edu}

\author{Aiden Prevey}
\affiliation{%
  \institution{University of Washington}
  \streetaddress{Guggenheim Hall}
  \city{Seattle}
  \state{WA}
  \country{USA}}
\email{aprevey@uw.edu}

\author{Carson Taylor}
\affiliation{%
  \institution{University of Washington}
  \streetaddress{Guggenheim Hall}
  \city{Seattle}
  \state{WA}
  \country{USA}}
\email{taylcar@uw.edu}

\author{Anika Tak}
\affiliation{%
  \institution{University of Washington}
  \streetaddress{Guggenheim Hall}
  \city{Seattle}
  \state{WA}
  \country{USA}}
\email{anikat9@uw.edu}

\author{Anthony Guinto}
\affiliation{%
  \institution{University of Washington}
  \streetaddress{Guggenheim Hall}
  \city{Seattle}
  \state{WA}
  \country{USA}}
\email{arguinto@uw.edu}

\author{Eeshani Shilamkar}
\affiliation{%
  \institution{University of Washington}
  \streetaddress{Guggenheim Hall}
  \city{Seattle}
  \state{WA}
  \country{USA}}
\email{eeshanis@uw.edu}

\author{Natarith Cheenaruenthong}
\affiliation{%
  \institution{University of Washington}
  \streetaddress{Guggenheim Hall}
  \city{Seattle}
  \state{WA}
  \country{USA}}
\email{natarith@uw.edu}

\author{Martin Nisser}
\affiliation{%
  \institution{University of Washington}
  \streetaddress{Guggenheim Hall}
  \city{Seattle}
  \state{WA}
  \country{USA}}
\email{nisser@uw.edu}


\renewcommand{\shortauthors}{Faruqi, Nisser}

\begin{abstract}

This paper introduces \system{}, a desktop wirebending machine and computational design tool for creating 3D wireframe structures. Combined, they allow users to rapidly and inexpensively create custom 3D wireframe structures from aluminum wire. Our design tool is implemented in freely available software and allows users to generate virtual wireframe designs and assess their fabricability. A path-planning procedure automatically converts the wireframe design into fabrication instructions for our machine while accounting for material elasticity and kinematic error sources. The custom machine costs \$293 in parts and can form aluminum wire into 3D wireframe structures through an ordered sequence of feed, bend, and rotate instructions. Our technical evaluation reveals our system's ability to overcome odometrically accumulating errors inherent to wirebending in order to produce accurate 3D structures from inexpensive hardware. Finally, we provide application examples demonstrating the design space enabled by Wirebend-kit.

\end{abstract}



\begin{teaserfigure}
  \includegraphics[width=\textwidth]{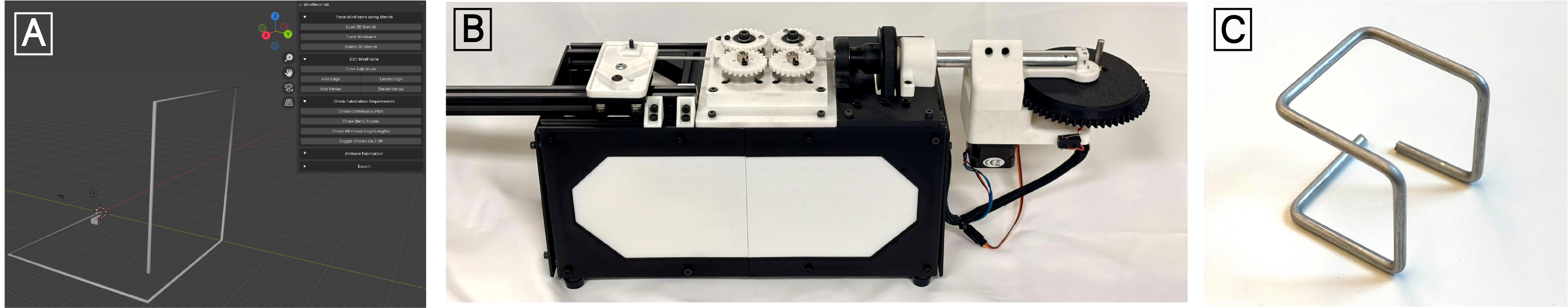}
  \caption{\system{} overview. (A) A computational design tool allows users to upload a 3D model as a stencil from which to create a wireframe design. The tool converts the design into wirebending instructions, a custom machine code consisting of feed, bend, and rotate commands which can be animated within the tool. (B) Once satisfied, instructions are uploaded to our custom wirebending machine to fabricate the object. (C) Our aluminum wireframe objects are strong, recyclable, and rapidly fabricated, opening new applications in personal fabrication and low-fidelity rapid prototyping.}
  \Description{Wirebending pipeline.}
  \label{fig:teaser}
\end{teaserfigure}

\maketitle

\section{Introduction}
Wirebending is a metal forming process used to bend wires, rods, or tube feedstock into wireframe structures. In its manual form, wirebending has been used to produce both functional and decorative artefacts such as jewelry since antiquity~\cite{oddy1977production}, and customized artefacts are still produced by craftspeople today. More recently, automated wirebenders developed in the 20th century operate primarily on an industrial scale, mass-producing household products that include coat hangers, springs and paperclips. However, combining this automation with customization can develop wirebending into a \textit{personal fabrication} platform that addresses key personal fabrication goals related to (1) materials, (2) sustainability, (3) speed, and (4) ease-of-use~\cite{baudisch2016personal}. 

Personal wirebending can \textit{(1) expand the range and quality of usable materials} in personal fabrication by processing aluminum and other metal feedstock that are strong, inexpensive, and ubiquitously available. It can also \textit{(2) foster sustainability in fabrication}. While 3D printers are the most popular personal fabrication technology with over 2 million sold by 2019~\cite{sargent20193d}, the most widely used materials in Fused Filament Fabrication, PLA and ABS, suffer from significant recycling challenges~\cite{rivera2023designing}. ABS is produced from non-renewable petroleum~\cite{pakkanen2017use}, while PLA is largely rejected from recycling facilities to end up in landfills due to stream contamination and loss of mechanical integrity~\cite{shen2011comparative}. In contrast, wirebent objects made from aluminum can be widely and repeatedly recycled without loss of mechanical integrity, cutting 95\% of the energy and emissions required for original processing ~\cite{das2010aluminum}. Personal wirebending can also \textit{(3) increase the speed and interactivity of fabrication}. Unlike additive manufacturing, which typically requires controlled processing to melt, sinter, or cure material layer-by-layer, wirebending is a forming technique. As such, wirebent shapes can be acquired rapidly via mechanical deformation alone, and can be reversibly deformed through continued interaction. While bending 1D wire offers speed advantages similar to folding 2D sheets~\cite{nisser2021laserfactory}, 2D folding typically requires pre-fabrication of planar faces~\cite{nisser2016feedback,niu2023pullupstructs} prior to folding.

The fourth goal requires wirebending tools to \textit{(4) overcome fabrication domain knowledge to increase ease-of-use}. Wirebending holds promise as a personal fabrication technology by utilizing strong and inexpensive materials, supporting sustainability, and affording rapid fabrication of low-fidelity wireframe objects. Despite these advantages, there are two key challenges that require addressing to overcome domain knowledge and increase their ease-of-use. First is a design tool that supports users to both design wireframe objects and generate their fabrication instructions. Previous work has explored methods to generate self-stabilizing~\cite{miguel2016computational} and articulated~\cite{liu2017wirefab} wireframes, or generate wireframes in augmented reality~\cite{feng2024yarmixedrealitycad}. However, tools that allow users to author custom 3D wireframes with feedback about their fabricability are limited. Second is the development of a wirebending machine in an inexpensive, desktop form factor that supports the automatic creation of custom wireframe objects without human intervention. Wirebending researchers and makers rely on expensive commercial machines~\cite{WireWare} or inadequately precise DIY tutorials to build their own~\cite{howtomechatronics,instructables}, significantly limiting access to practical wirebending applications. 




In this paper, we introduce a desktop wirebending machine and a computational design tool to rapidly create custom 3D wireframe structures from aluminum feedstock. In the design tool, users upload a 3D model from which to generate wirebending instructions that can be both animated within the tool and fabricated on the machine. The tool guides users to design a fabricable wireframe, allows animating the full fabrication sequence to check for collisions, and converts the wireframe into fabrication instructions for the machine. Our custom machine costs \$293 in parts and fully automates fabrication to produce cm-scale wireframe objects in approximately 3-9 minutes. We designed and built the machine to maximize the use of commercial electronic components and 3D-printed parts to ease assembly and minimize cost for users. A key challenge to wirebending accurate 3D geometries is that errors accumulate odometrically throughout the fabricated part. To address this, we develop an error model by characterizing material and kinematic error sources and validate this experimentally, allowing software to de-bias errors arising from inexpensive hardware and thereby reducing reliance on more accurate but expensive electromechanical components. Finally, we use \system{} to design and fabricate wireframe objects that demonstrate the speed and versatility of our system. All objects fabricated by our system during development were recycled by our institution's Facilities department. 

\section{Related Work}

Our work is primarily related to research in computational fabrication of wireframe structures, both with and without wirebending. We first survey work addressing the creation of wireframe structures, then position our work in relation to design tools and fabrication machines for the wirebending process.

\subsection{Wireframe structures} 

Wireframe objects offer high strength-to-volume ratios, rapid fabrication times, and support users to evaluate key affordances of an objects' scale and ergonomics. Seizing on these advantages, researchers have explored how to design and fabricate wireframe structures using a variety of fabrication techniques.


Using 3D printing for wireframe objects, \textit{WirePrint}~\cite{mueller2014wireprint} and \textit{On-The-Fly-Print}~\cite{peng2016fly} used FDM 3D printers to rapidly print certain wireframe structures, providing a low fidelity preview to full-resolution 3D prints. FrameFab~\cite{huang2016framefab} show how thermoplastic extruders can be deployed to 3D print general frame shapes. In related work, \textit{WireDraw}~\cite{yue2017wiredraw} used a 3D extruder pen together with a mixed reality system to guide direct drawing of 3D wire sculptures. Trussfab~\cite{kovacs2017trussfab} allowed users to create large load-bearing truss structures that guides users through manually assembling plastic bottles connected by 3D-printed attachments. 


Others have explored creating wireframes that can be \textit{manually} bent. \textit{WrapIt}~\cite{iarussi2015wrapit} developed a design tool for creating 2D wire-wrapped jewelry which can be manually bent around 3D-printed jigs. Extending this to 3D, Tojo et al.~\cite{tojo2024fabricable} and Wang et al.\cite{10.1145/3355056.3364575} generate 3D-printed jigs to manually bend wire to form 3D wire art. \textit{WireRoom}~\cite{yang2021wireroom} introduce a design tool to create minimalistic 3D wire art which is manually bent by participants in 15-45 minutes.



While several modalities have been demonstrated to design wireframe objects, supporting the fabricability of these designs is an outstanding problem, particularly without relying on manual bending, which is a challenging and unintuitive task. In this work, we introduce a tool that supports users to author designs, assess their fabricability, and generate fabrication instructions that can be automatically executed by a custom wirebending machine. 



\begin{figure*}[t]
  \centering
  \includegraphics[width=0.99\linewidth]
{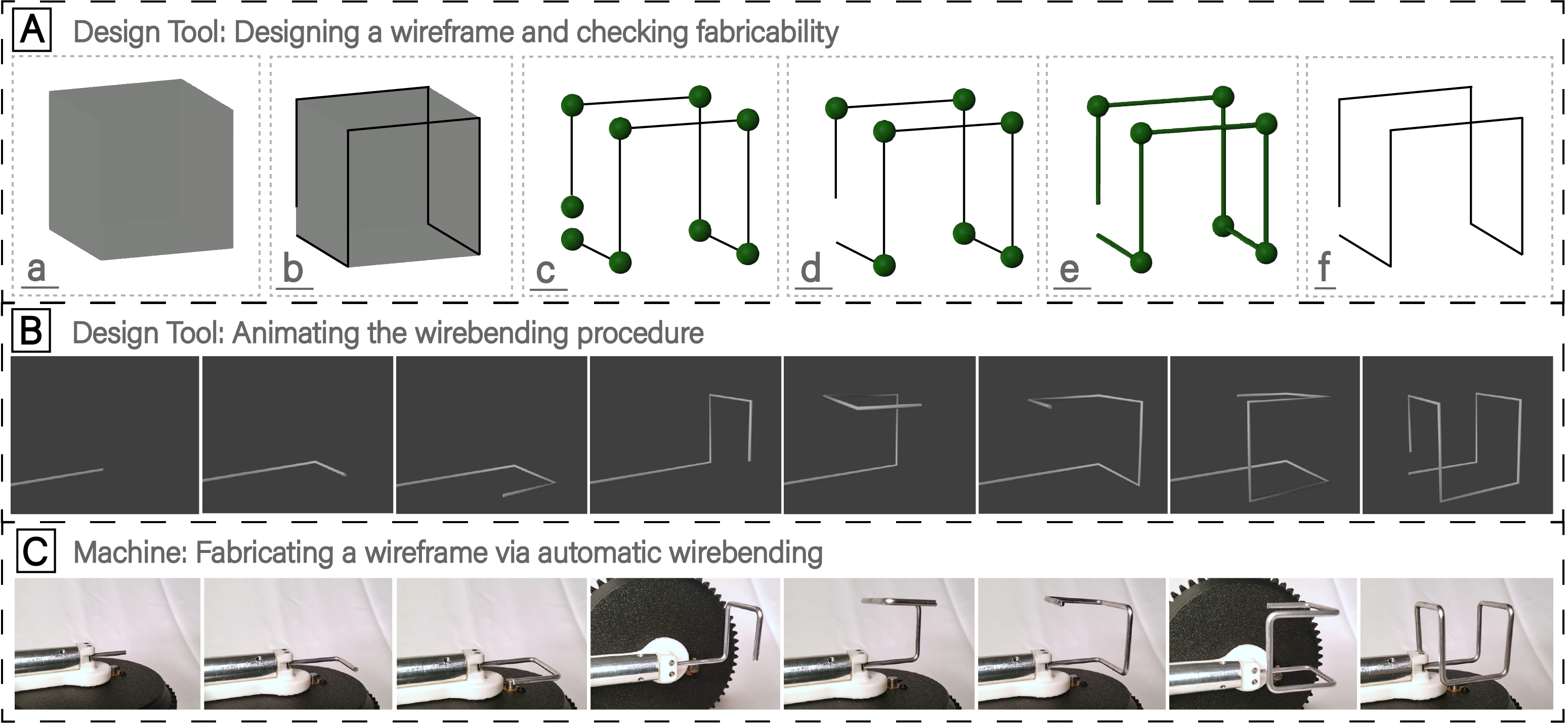}
  \caption{Key steps to using \system{}: designing, animating, and fabricating a wireframe. (A) In the design tool, users can (A-a) upload a 3D stencil from which to (A-b) trace a wireframe design and (A-cde) check fabricability. Vertices or edges are highlighted in green if (A-c) a continuous path exists, (A-d) bend angles are within machine limits, and (A-e) feed lengths are within machine limits. (A-f) Once checks are passed, fabrication instructions are generated. (B) An animation allows users to inspect every step in the fabrication procedure before (C) uploading instructions to the machine for fabrication.}
  \Description{system-overview}
  \label{fig:system-overview}
\end{figure*}

\subsection{Wirebending design tools}

Software for wirebending machines~\cite{bendtech,VGPNext,WireWare} exist as closed-source commercial products associated to particular commercial machines, but often require defining sequential low-level machine operations to generate instructions for complex shapes. To allow users to prescribe high-level wireframe designs, researchers have developed tools that support both designing and evaluating the physical fabricability of these designs on wirebending machines.

\textit{Y-AR}~\cite{feng2024yarmixedrealitycad} introduced a mixed reality CAD interface to wirebend 3D wireframes that connect to other objects, using heuristics to support fabricability. \textit{WireShape}~\cite{gohlke2023wireshape} translate 2D vector outlines into instructions for a 2D wirebender and warns users if certain fabricability constraints are violated, requiring users to refine wirebent structures manually using tools and bending jigs.
\textit{WireFab}~\cite{liu2017wirefab} developed a design pipeline with mixed fabrication processes to create articulated structures by connecting 3D wirebent objects. In related work, \textit{Bend-It}~\cite{xu2018bend} introduce a technique for physically animating wirebent 2D structures using cable-drive attachments.

Other research focuses on generating collision-free wirebending paths that are fabricable. Baraldo et al.~\cite{baraldo2022automatic} developed an algorithm based on A* to find collision-free bending sequences given a target geometry and machine architecture, but were limited to sequences under 8 bends. Miguel et al.~\cite{miguel2016computational} support users to bend 2D shapes on a commercial 2D machine and manually assemble these to create interlocking 3D structures with stability. Lira et al.~\cite{lira2018fabricable} generate a small number of machine-fabricable subwires that can be tied together to produce a wireframe 3D structure. \textit{Bend-forming}~\cite{bhundiya2023bend} develop a path-planning technique to create complex reticulated structures using a commercial wire-bending machine, modeling the residual stress in fabricated structures and securing final geometries using mechanical joints that connect disjoint vertices. \textit{Tune-it}~\cite{wu2024tune} generate an intermediate collision-free path for machine bending which is then manually corrected by a user, also supporting circular segments, however fabricated examples are limited to 2D shapes.

In this work, we build on prior research in three key ways. First, we create a design tool that allows users to author 3D wireframes by annotating 3D stencils\textemdash an uploaded 3D model. Unlike previous work that automatically extracts paths~\cite{liu2017wirefab}, we cede design authority to the designer and support them with fabricability checks for their intended designs. This strategy draws on personal fabrication research that leverage stencils to design 3D objects that are printable~\cite{savage2015makers}, augmented with electronics~\cite{10.1145/3196709.3196791} or rendered for inspection~\cite{lau2010modeling}. Second, our design tool includes an error model that permits accurate wirebending of 3D designs without manual manipulation or costly electronics. Our model builds on previous work~\cite{bhundiya2023bend} to incorporate a highly significant 'setback' term which is key for wider diameter feedstock. This is crucial because errors in sequentially bent or folded structures accumulate in the structure. Finally, the tool alerts users if a design violates machine constraints on edge length, bend angle, and path existence, and allows simulating the wirebending process for users to inspect for collisions. 

\subsection{Wirebending machines}

Wirebending machines come in 2D and 3D variants, and are capable of deforming metal wire into 2D or 3D shapes, respectively, through an ordered sequence of feed, bend, and rotate instructions. The machines used in prior research typically fall into two categories. One group of studies~\cite{bhundiya2023bend,feng2024yarmixedrealitycad,miguel2016computational,xu2018bend,baraldo2022automatic} leverage \textit{commercial} machines, most popularly Pensa~\cite{WireWare}. Commercial wirebenders can be expensive; 3D machines from one supplier cost \$68500, and one 2D wirebender is quoted~\cite{xu2018bend} to cost \$3675. Compounding cost constraints, these researchers note that closed-source commercial machines exhibit limitations for research due to lack of programmability. For these reasons, other studies~\cite{liu2017wirefab,lira2018fabricable,gohlke2023wireshape} build their own wirebending machines, typically using online DIY tutorials~\cite{instructables,howtomechatronics}. These tutorials typically include a bill of materials, CAD files, and arduino code for executing key commands. However, these DIY architectures are not thoroughly documented, introduce unnecessary complexity via wire straightening mechanisms that fail to eliminate wire kinks, exclude components necessary to prevent fatiguing and error drift, and exclude error-correction software necessary for accurate wirebending, leading to angle errors that can exceed 15\degree{}. Other studies design new machines, but cannot be recreated for personal fabrication because architectures use only machined parts or lack the documentation required to create and operate them~\cite{farhan2024design,hamid2018integration,von2024reducing}.

To address these challenges, we develop a custom 3D wirebending machine for a parts cost of \$293 and document our design. We optimize our machine to rely on 3D printed parts to support ease-of-assembly. In particular, we show how our machine architecture and error-correction model reduce errors feed and bend angles by up to 17x compared to model-free fabrication, supporting the personal fabrication of highly customized objects.

\begin{figure*}[ht!]
  \centering
  \includegraphics[width=0.9\linewidth]
{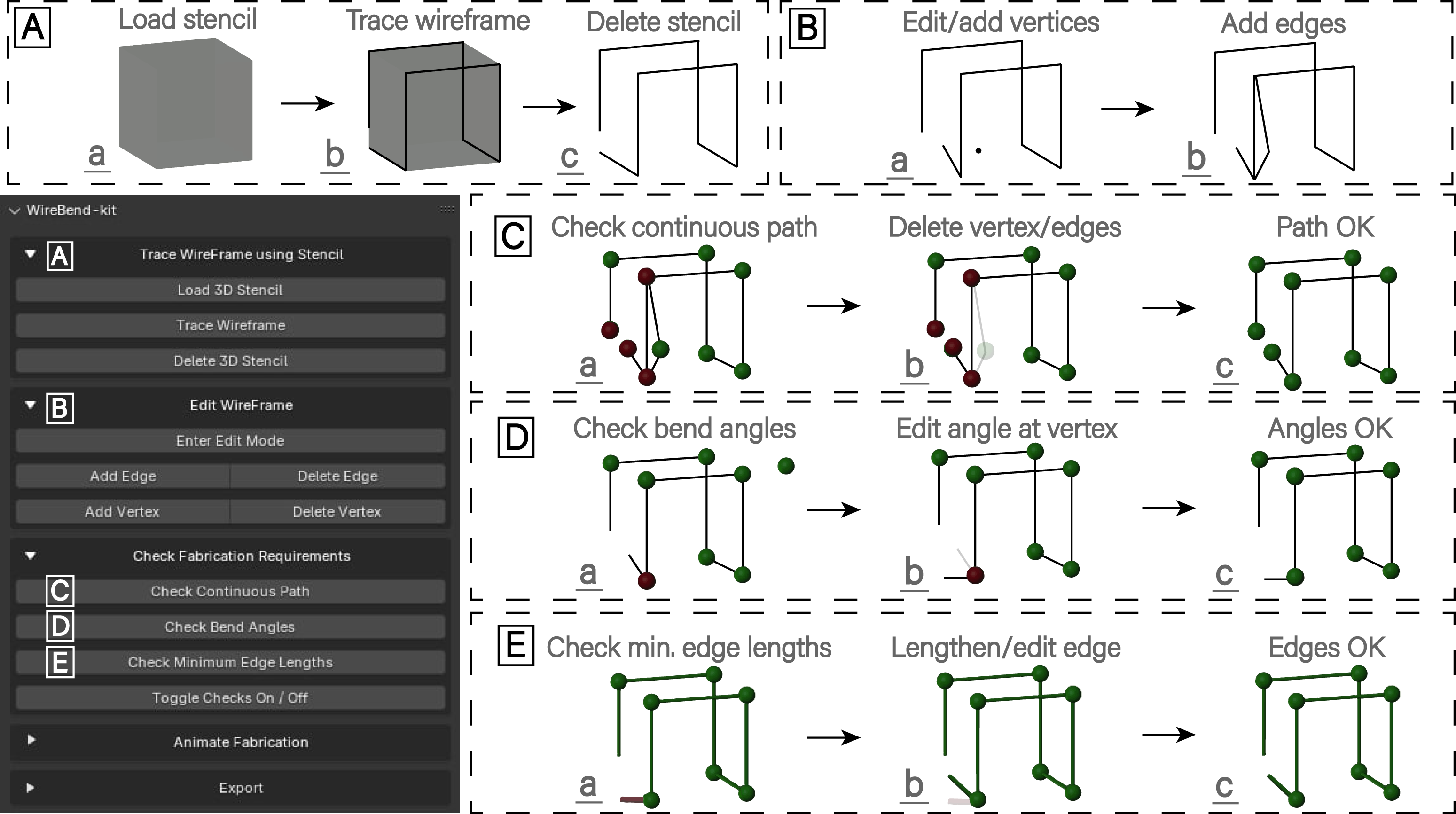}
  \caption{ User interface and workflow for interactive wireframe design and fabrication validation. (A) Users import a 3D stencil and trace a wireframe over it, which can then be edited. (B) The traced wireframe can be modified by adding or removing vertices and edges. (C–E) The system supports real-time fabrication checks to ensure the design satisfies constraints: (C) Eulerian continuity for uninterrupted wire bending, (D) maximum allowable bend angles, and (E) minimum segment length. Issues are highlighted in red and resolved through interactive editing, with updates reflected immediately in the 3D viewport.}
  \Description{UI}
  \label{fig:UI}
\end{figure*}

\section{Overview of \system}

\system{} consists of two main parts: a computational design tool that allows users to design wireframes, and a wirebending machine that fabricates them. 

In the design tool, users upload a 3D model as a stencil from which to create a wireframe design (Figure \ref{fig:system-overview}A). The interface provides feedback to the user to help ensure the design meets key fabrication constraints. In particular, the system evaluates whether a continuous path required for wirebending exists, whether maximum permissible bend angles are violated, and whether minimum edge lengths are violated. If violations occur, these are highlighted in red; if checks are passed, they are highlighted in green. The interface allows users to move, add, and delete vertices and edges to satisfy these constraints or to make and edit designs.  

Once a design is finalized, the tool converts the design into wirebending instructions; a custom machine code consisting of feed, bend, and rotate commands. These instructions are automatically post-processed with an error-adjustment model that accounts for material and kinematic error sources resulting from the machine architecture and materials. Our model is symbolic, and can be updated if materials or the machine architecture change. Using the error-adjusted instructions, users can animate every step in the fabrication process to check for potential self-intersections and inspect the model prior to fabrication (Figure \ref{fig:system-overview}B).

Once satisfied, users export the instructions to a graphical user interface (GUI) which can command the machine to fabricate the part (Figure \ref{fig:system-overview}C). The GUI also allows users to manually execute sequential feed, bend and rotate instructions to create free-form parts, which can be saved and replayed at a later date.

\section{Computational Design tool}





Our computational design tool helps users translate abstract 3D models into fabricable wireframe structures. The tool supports the design process with real-time fabricability checks, animates the wirebending process for user inspection, and generates error-corrected fabrication instructions which are exported to a GUI that commands the machine.

\subsection{User interface}
\label{sec:user_interface}
We developed a custom Blender plugin for tracing, editing, and analyzing 3D wire structures (Figure~\ref{fig:UI}). Integrated into Blender’s native UI as a sidebar panel, it includes 5 collapsible sections: (1) Trace Wireframe Using Stencil, (2) Edit Wireframe, (3) Check Fabrication Requirements, (4) Animate Fabrication, and (5) Export. To enable construction and fabrication-aware editing of wireframe structure with our UI tool, we demonstrate construction of the simple cube from Figure~\ref{fig:system-overview}. By first modifying the cube to violate fabrication constraints, we show how our tool can aid users in identifying and resolving fabrication issues before fabricating the part.  

\subsubsection{Creating Wireframes with Stencils}
Users start by opening the plugin from the sidebar and opening the Trace Wireframe section (Figure~\ref{fig:UI}A). The user imports a reference 3D model (the 'stencil'), using the 'Load 3D Stencil' button, which renders the model as semi-transparent to facilitate visual tracing. Users enable 'trace mode' by clicking the 'Trace Wireframe' button, and generate a wireframe by clicking on the stencil's surface to place vertices which snap to the closest vertex on the stencil. As the user continues to click on vertices, subsequent vertices are connected automatically with edges to form the wireframe trace. 

\subsubsection{Editing WireFrame}
Once an initial design is created, users can hide the stencil by clicking 'delete 3D stencil'. Next, the 'Edit Wireframe' section (Figure~\ref{fig:UI}B) allows users to refine the traced structure. They can add or remove vertices and edges through custom buttons embedded in the UI. We demonstrate this by adding another vertex to the exemplar cube, by adding a vertex and two edges. The combined stencil and free-form modification technique allows users to first create a base design, then refine it based on their desired aesthetic or constraint feedback. 

\subsubsection{Checking for Fabrication Requirements}
\label{sec:ui_check_fabrication_requirements}
To ensure that user-designed wireframes are physically realizable with our wirebending machine, our system automatically checks for three key fabrication constraints (Figure~\ref{fig:UI}C-E): (1) Eulerian continuity, (2) maximum allowable bend angles, and (3) minimum edge length. These constraints reflect constraints imposed by the wirebending hardware, and the UI provides real-time feedback about constraint violations during modeling. Each constraint is surfaced through the section titled 'Check Fabrication Requirements'. Users can trigger validation routines that analyze the wireframe and apply visual annotations to the 3D model. These annotations appear as colored spheres where vertices are checked for Eulericity or subtended bend angles, and as colored tubes where edges are checked for length constraints. A Toggle Annotations button will hide or reveal these indicators. This enables iterative design where users can continuously refine their geometry while receiving real-time guidance.


Wirebending machines bend a single continuous wire, requiring the bending path to be Eulerian. This requires every vertex to either have even degree, or for exactly two vertices to be of odd degree. Our system represents the wireframe mesh as a graph using the \texttt{NetworkX} library, mapping vertices to graph nodes and edges to graph connections. It then evaluates the degree of each node and annotates vertices using colored spheres in the 3D viewport. If the graph contains two or zero odd-degree vertices, markers appear green to indicate a valid fabrication path. If there are one or more than two odd-degree vertices, such as in Figure~\ref{fig:UI}C, all odd vertices are highlighted in red, prompting the user to make design changes. 

Wirebending machines exhibit a maximum bend angle; how sharply a wire can be bent before the wire collides with the machine or itself. To address this, our system calculates the subtended angle at each vertex based on adjacent edges. If the angle exceeds the machine’s configured range ($\pm155\degree$), a red warning marker is attached to the vertex as in Figure~\ref{fig:UI}D. The user can then edit the vertex location or an adjoining edge to reduce this angle. The vertex color switches to green once the subtended angle is within the viable range. This feature ensures that wireframes remain fabricable and within safe geometric tolerances.

Wire segments require a minimum length defined by the bending radius of the machine's bending peg. Our system checks every edge's Euclidean length and annotates in red (Figure~\ref{fig:UI}E) those shorter than a predefined threshold of 20.4mm. This switches to green when the user edits the edge to satisfy the constraint.

\begin{figure}[t]
  \centering
  \includegraphics[width=0.95\linewidth]
{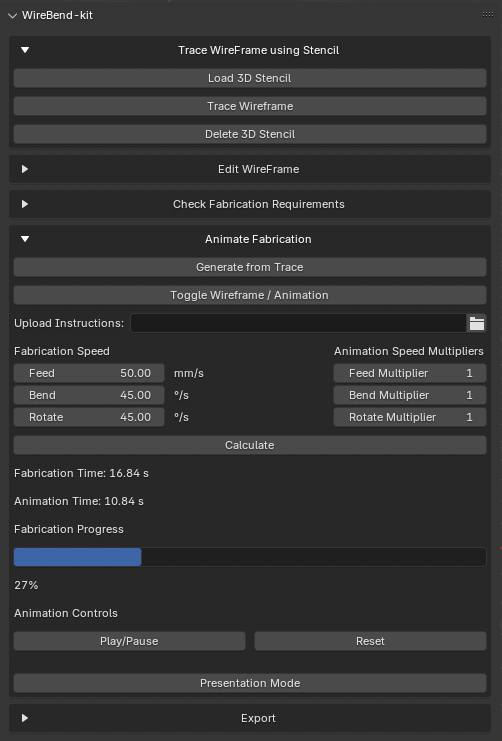}
  \caption{WireBend-kit UI for animating the fabrication process. Users can specify fabrication parameters—including feed, bend, and rotation speeds—and adjust animation playback through speed multipliers. The estimated fabrication time is displayed, and a progress bar allows monitoring the simulated fabrication in real time. Controls for playing, pausing, and resetting the animation facilitate iterative exploration and validation of the bending sequence.}
  \Description{animation}
  \label{fig:animation}
\end{figure}

\subsubsection{Animate Wireframe}

To support previewing the fabrication process, our UI contains a 'Animate Fabrication' section that animates the full wirebending procedure (Figure~\ref{fig:animation}). This animation module converts wireframe design into a step-by-step visualization of how the wire is sequentially fed, bent, and rotated. 


Users trigger the animation by clicking 'Generate from Trace' which computes the traversal path using Hierholzer’s algorithm and calculates feed distances, bend angles, and rotation angles. Next, users can configure fabrication-specific parameters such as feed speed (mm/s), bend speed (degrees/s), and rotation speed (degrees/s) to calculate the true fabrication time of the part using up-to-date machine speeds. These speeds can be sped up or down independently inside the animation via speed multipliers to slow or accelerate specific fabrication stages for better visibility. Clicking 'Play/Pause' will simulate the entire fabrication sequence, feeding, bending and rotating the wire to reflect the physical counterpart. A progress bar and percentage indicator show the completion rate. Users can pause and resume the animation at any point, or reset the animation using the "Reset" button, restoring the original wireframe and clearing all bend highlights. This real-time animation helps users validate that paths are collision-free and can rectify any conflicts by re-editing the wireframe. 

\subsubsection{Exporting for Fabrication}
Once the user is satisfied by the animation, the 3D wireframe can be exported for future use, and the fabrication instructions can be exported as a \texttt{txt} file for the machine to fabricate. On export, the instructions file is post-processed using an error correction model detailed in section \ref{sec:error-corr}.

\section{From Wireframe to Instructions}
Our UI allows users to rapidly design wireframes and convert these into fabricable instructions that satisfy fabrication constraints. In this section, we describe the key processes that are automated for the user in order to ease design: confirming the presence of an euler path, translating this path into machine instructions, and post-processing instructions for error-correction. 

\subsection{From Mesh to Bending Path}
Once a user has finalized a wireframe design using \system{}, we parse the mesh to extract all vertices and their 3D coordinates, and construct an adjacency matrix from the edges. Since the mesh represents an undirected wireframe, the adjacency is symmetric. This procedure ensures that every edge present in the geometry is captured in the matrix. Using the adjacency information, we instantiate a graph $G = (V, E)$ where each vertex in \(V\) corresponds to a mesh vertex annotated with its 3D coordinates. To avoid redundant edge representations in the undirected graph, we add each edge only once (i.e., for vertex pairs \((i, j)\) with \(i < j\)). Additionally, we assign a unique edge identifier (\texttt{edge\_id}) to each edge. 

We then ensure that the underlying graph structure is \textit{Eulerian}—that is, it contains a path that visits every edge exactly once. This is essential for producing an admissible wirebending sequence from a continuous length of wire. A graph is Eulerian if and only if all its vertices have even degree and the graph is connected. If exactly two vertices have odd degree, the graph admits an \textit{Eulerian trail} (open path); otherwise, no such path exists. We verify whether the graph is Eulerian by examining the degree of each vertex. If all nodes have even degree, a closed Eulerian circuit exists and we proceed directly to path extraction using Hierholzer’s algorithm~\cite{hierholzer1873moglichkeit}. We do not employ automatic methods to Eulerize user-defined wireframes, as an automatically modified wireframe may deviate from the author's design intent. Instead, we highlight any problematic vertices in the UI for the the user to edit (Section~\ref{sec:ui_check_fabrication_requirements}).







\subsection{Graph to Fabrication Instructions}

\begin{figure}[h]
  \centering
  \includegraphics[width=0.99\linewidth]
{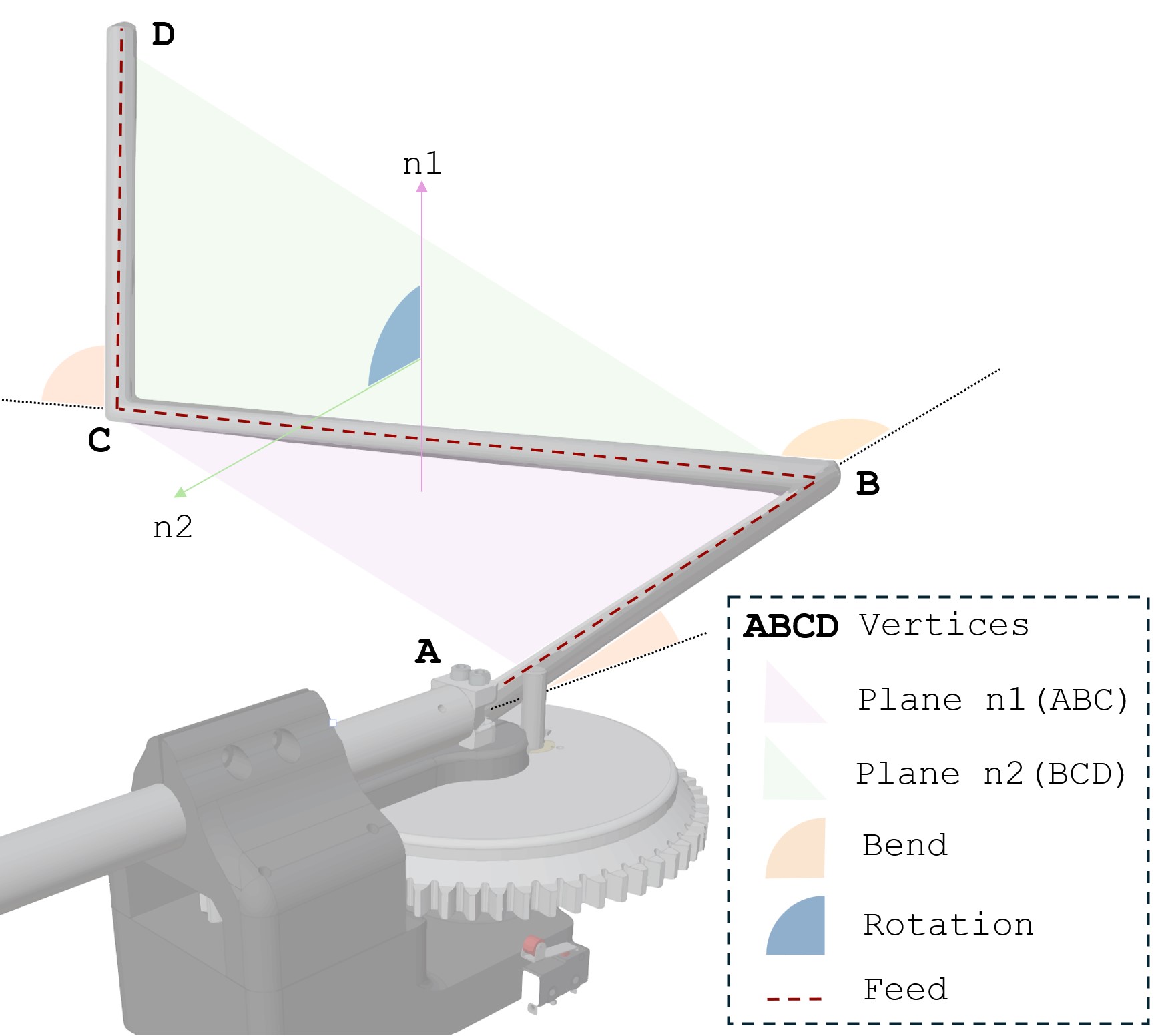}
\caption{Creating wirebending instructions from a graph. A Eulerian path (A→B→C→D) is converted into sequential machine instructions. Feed lengths are computed between nodes, bend angles from node triplets (e.g., $\angle$ABC), and rotate angles from the normals of adjacent bend planes (e.g., $\vec{n}_1$, $\vec{n}_2$).}

  \Description{graph-instructions}
  \label{fig:graph-instructions}
\end{figure}

Once an Euler path is found, we convert the path into machine-executable wirebending instructions (Figure \ref{fig:graph-instructions}): feed lengths, bending angles, and rotation angles. While feed lengths correspond simply to the euclidean distance between consecutive vertices, the angular components of the path must be decomposed into two distinct operations. Specifically, a \textit{bend angle} represents the angle subtended between two neighboring edges, and is physically bent on the machine. In contrast, a \textit{rotation angle} is computed to rotate the bending mechanism into a new plane, and does not itself bend the wire. Combined, these allow producing 3D wireframes using planar bending operations interleaved with rotations. Instructions are finally outputted as \texttt{F} (feed), \texttt{B} (bend), and \texttt{R} (rotate) commands.

\paragraph{Feed Length (\texttt{F})}
For each pair of consecutive nodes \( \mathbf{v}_i, \mathbf{v}_{i+1} \) in the path, we compute the Euclidean distance between them \( \mathbf{F}_i\) to determine the amount of wire to feed forward. This is output as an \texttt{F} instruction: \texttt{F} $F_i$.

\paragraph{Bending Angle (\texttt{B})}
At each intermediate node \( \mathbf{v}_{i} \), we compute the angle \( \theta_i \in [0, \pi] \) between the incoming vector \( \mathbf{v}_i - \mathbf{v}_{i-1} \) and the outgoing vector \( \mathbf{v}_{i+1} - \mathbf{v}_i \). This is output as a \texttt{B} instruction: \texttt{B} $\theta_i$.

\paragraph{Rotating Angle (\texttt{R})}
Three consecutive but non-collinear points can always be used to define a unique bending plane in 3D space. However, as additional vertices are introduced, consecutive bends may no longer lie in the same plane, and will require rotating the bending plane. This requires computing a rotation angle between two adjacent bending planes: one defined by the triple \((\mathbf{v}_{i-2}, \mathbf{v}_{i-1}, \mathbf{v}_i)\), and the next by \((\mathbf{v}_{i-1}, \mathbf{v}_i, \mathbf{v}_{i+1})\). We find the normals to these planes and compute the signed rotation angle between them as an \texttt{R} instruction: \texttt{R} $R_i$. If either normal vector is undefined (i.e., due to collinearity), the rotation is set to zero.

The full instruction sequence is returned as a list of interleaved commands
($\texttt{F}~F_i , \texttt{R}~R_i , \texttt{B}~B_i$) that are written to a file for animation in the UI and fabrication on the machine.

\begin{figure*}[t]
  \centering
  \includegraphics[width=0.9\linewidth]
{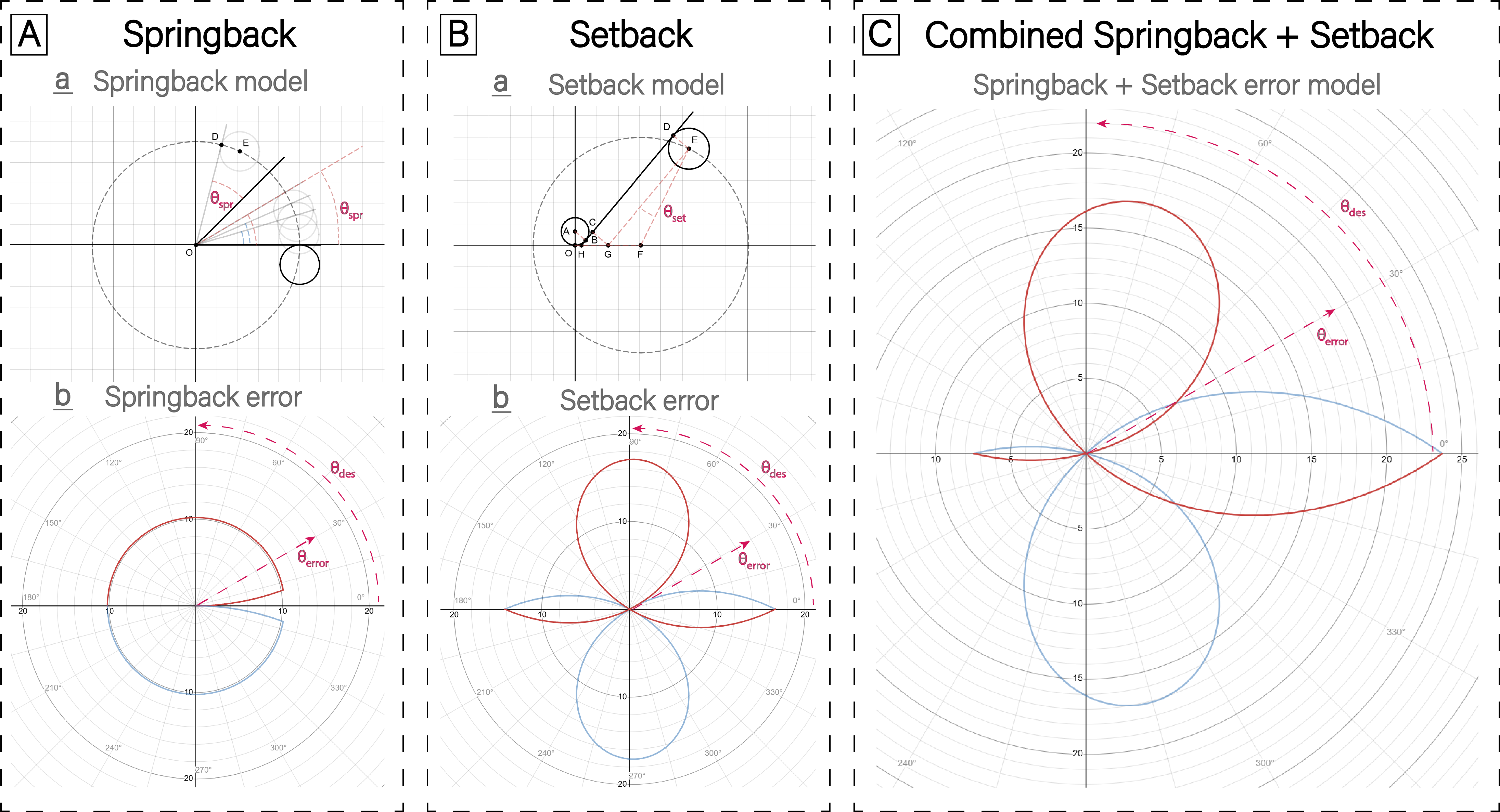}
  \caption{Modeling and compensating for bending errors: (A) springback, (B) setback, and (C) combined models. (A-a) A geometric model of springback is used to derive a (A-b) springback error-compensation model to address the tendency of aluminum to elastically "spring back" after being bent. (B-a) A geometric model of setback is used to derive a (B-b) setback error-compensation model to address the issue that the wire does not bend to the same angle that the bending peg sweeps. (C) We combine these error models to adjust the bend angles commanded to our system.}
  \Description{Error model}
  \label{fig:BendError}
\end{figure*}

\subsection{Error correction}
\label{sec:error-corr}
The instructions generated by our tool represent the true edge lengths, bend angles, and rotation angles between sequential edges. However, fabricating a physical object that is true to size requires error compensation that adjusts these values before fabrication.

\subsubsection{Error correction for Feed}

When wire is bent on a wirebending machine, the material strains plastically\textemdash lengthening\textemdash at the bend. Without adjustment, this can significantly impact the size of fabricated models, producing edge lengths that are up to 10\% greater than designed edge lengths for cm-scale wireframes.

We develop a model based on sheet metal bend deduction to adjust feed lengths through a geometric adjustment that accounts for the material strain when bending wire from a straight line into an arc. This arc length can be approximated using the bend angle and an arc-length formula, using a radius defined from the nozzle rod to the neutral axis of the wire. The neutral axis is used as no strain occurs at this point in the cross section, and is found using a material's K-value and diameter. For aluminum, typical K-factors of approximately 0.45 to 0.50 are commonly used, however our empirical testing suggests a K-value closer to 0.3 for our stock. For an edge length in between two bends, it is necessary to account for strain-lengthening at both of these bends, while also accounting for "lost" length due to arc curvature. The final approximation of the adjusted feed length ($F_{adj}$) subtracts the lost path distance ($l_{lost}$) for both bends, adds the arc length ($l_{arc}$) of the later bend, and accounts for the offset between the center of the wire ($r_{wire}$) and the neutral axis ($c$). Our model for this adjustment is given below:

\begin{equation}    
F_{adj}\left(n\right)=F_{des,n}-l_{lost}\left(\theta_n-\theta_{n+1}\right)+l_{arc}\left(\theta_{n+1}\right)-2\left(r_{wire}-c\right)
\end{equation}

\subsubsection{Error correction for Bend}

Our bend error model accounts for two main sources of error: springback and setback (Figure \ref{fig:BendError}). 

\paragraph{Springback:} In elastic materials such as aluminum, springback refers to their tendency to "spring back" to smaller strains, or smaller bend angles. Figure \ref{fig:BendError}A-a illustrates this concept, where a bending peg with centerpoint E rotates counter-clockwise to contact the wire at D, and O is the point from which the wire exits the nozzle and the peg rotates around. In this case, the peg has bent the wire from the X-axis to D, but the wire "springs back" clockwise by $\Theta_{spr}$. If the wire is bent counter-clockwise from the X-axis to an angle smaller than $\Theta_{spr}$, no permanent bend will occur. Figure \ref{fig:BendError}A-b depicts the resulting error plot in polar co-ordinates. Here, sweeping a desired positive bend angle $\Theta_{des}$ from the X-axis yields an angle error $\Theta_{error}$ given by the radius of the curve. Figure \ref{fig:BendError}A-b illustrates that our system shows a springback of 10.23\degree. 

\paragraph{Setback:} Setback is an unintuitive geometric error that arises because the point at which the wire bends is a function of the bend angle, and is affected by the distance between the bending nozzle and center of rotation of the bending mechanism. Figure \ref{fig:BendError}B-a illustrates this model. Once again, the bending peg with centerpoint E rotates counter-clockwise to contact the wire at D, and O is the nozzle exit. Here, we also depict the nozzle peg centered at A, which the wire contacts at B as it bends around it. However, point B changes as a function of bend angle, which affects the bent wire angle (vector BD). In addition, the center of rotation of the bending peg is at point F, meaning the angle swept by our bending peg is defined by vector FE. Accordingly, the angle swept by the bending peg (vector FE) is not parallel to the angle swept by the wire (vector BD), and this setback angle $\Theta_{set}$ is a non-linear function of the bend angle. Figure \ref{fig:BendError}B-b depicts the non-linear error $\Theta_{error}$ between the desired wire angle $\Theta_{des}$ and the bending peg angle.

\paragraph{Combined Springback-Setback model:} Using our models for springback and setback error, we derive a combined model of our bending error, shown in Figure \ref{fig:BendError}C. This model is symbolic in the geometry of the bending operation, and can be used for other nozzle geometries and machine configurations if the machine is re-designed. In this figure, we plot the combined model error for a distance between the nozzle exit (O) and bending center (F) of 12mm.  Below, we describe this model in greater detail.

The geometry for a given setback model forms the polygon (ABDEFO) shown in Figure \ref{fig:BendError}B-a. The setback model determines the commanded angle required for the machine $\theta_{com}$ ($\angle EFX_+$), where $X_+$ is the positive x-axis direction, given a desired wire angle $\theta_{des}$ ($\angle DHF$). The model was derived by dividing the polygon into triangles GEF and HCG and right kite ABHO. The resultant solution for the setback angle, $\theta_{set}$, as a function of the desired wire angle, $\theta_{des}$, can be written as:

\begin{equation}
\theta_{set} = \theta_{des} + \sin^{-1}(\frac{R}{s - \frac{r_r}{sin(\theta_{des})} - r_n\tan(\theta_{des})}\sin(\theta_{des}))
\label{eq:geometry}
\end{equation}

where, $R$ is the radius ($\overline{EF}$) of the arc traced by the bending rod, $r_r$ is the radius ($\overline{DE}$) of the bending rod, $r_n$ is the radius ($\overline{AB}$) of the nozzle, and $s$ is the setback distance $\overline{OF}$.

We define springback as a constant $S$ in the bending direction, which we determine by bending a series of angles after setback error is compensated for.

\begin{equation}
\theta_{spr} = \theta_{des} + \text{sgn}(\theta_{des}) S
\label{eq:geometry}
\end{equation}

The combined model for commanded angle, $\theta_{com}$, as a function of desired wire angle, $\theta_{des}$, is the setback model, $\theta_{set}$, with the springback model, $\theta_{spr}$, substituted in as the desired wire angle $\theta_{des}$ in the setback model:


\begin{equation}
\theta_{com} = \theta_{spr} + \sin^{-1}(\frac{R}{s - \frac{r_r}{sin(\theta_{spr})} - r_n\tan(\theta_{spr})}\sin(\theta_{spr}))
\label{eq:geometry}
\end{equation}

We use this combined model to translate desired bend angles $\theta_{des}$ from our UI into commanded angles $\theta_{com}$ sent to the machine.


\section{Using the machine}

\begin{figure}[h]
  \centering
  \includegraphics[width=0.99\linewidth]
{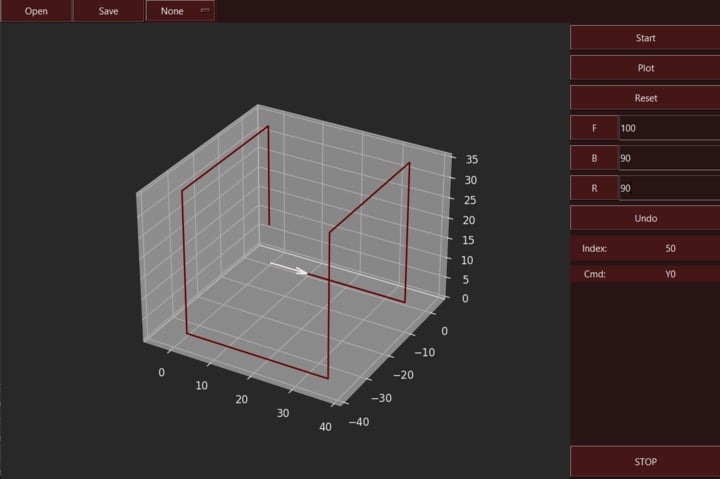}
  \caption{Users export fabrication instructions to a GUI that instructs the machine to execute fabrication. Users can also execute feed, bend and rotate commands manually.}
  \Description{GUI}
  \label{fig:GUI}
\end{figure}

After a set of machine instructions have been generated by the design-tool and automatically adjusted for error-correction, a python script launches a GUI to operate the machine (Figure \ref{fig:GUI}). The GUI communicates with the machine using the Pyserial library via a USB cable connected to the wirebender's Arduino Uno.

The GUI allows has two modes of operation: automatic and manual. In automatic operation, instruction files generated by the design tool are opened using the "open" button and animated within the GUI using the "plot" button. The "Start" button initiates fabrication and the "STOP" button will terminate any UI or fabrication processes. In manual operation, users may create wireframe objects without a pre-existing instruction file by clicking buttons F, B, and R, to execute feed, bend, and rotate commands by the argument inserted in the adjacent input fields. The animation displays the resulting wireframe digitally, and users can click the "save" button to store manually created objects to replay later. In both operation modes, the GUI translates the feed, bend and rotate commands into low-level machine instructions to operate the stepper motors while accounting for all gear ratios. On startup, a homing procedure also calibrates the bending motor to the angle that corresponds to 0\degree{}.

\section{Wirebending machine}

Our wirebending machine was designed to maximize the use of commercial electronic components and 3D-printed parts, in order to ease assembly and minimize costs. It costs \$293 in parts, including retail prices for commercial-off-the-shelf components and material costs for all custom 3D-printed and machined parts. Three parts are machined, and unless otherwise noted, all parts are 3D-printed using PLA on a Bambu X1C 3D printer. Costs for custom parts were generated by multiplying the material mass of our parts by the cost/kg of the bulk materials purchased. A complete list of parts, costs, purchasing links and CAD files will be published on our website. The machine fabricates wireframe objects from 3mm diameter feedstock of Aluminum 6061 T6, an inexpensive, common, and recyclable material which we purchase for 0.6\$/ft.

\begin{figure*}[h]
  \centering
  \includegraphics[width=0.9\linewidth]
{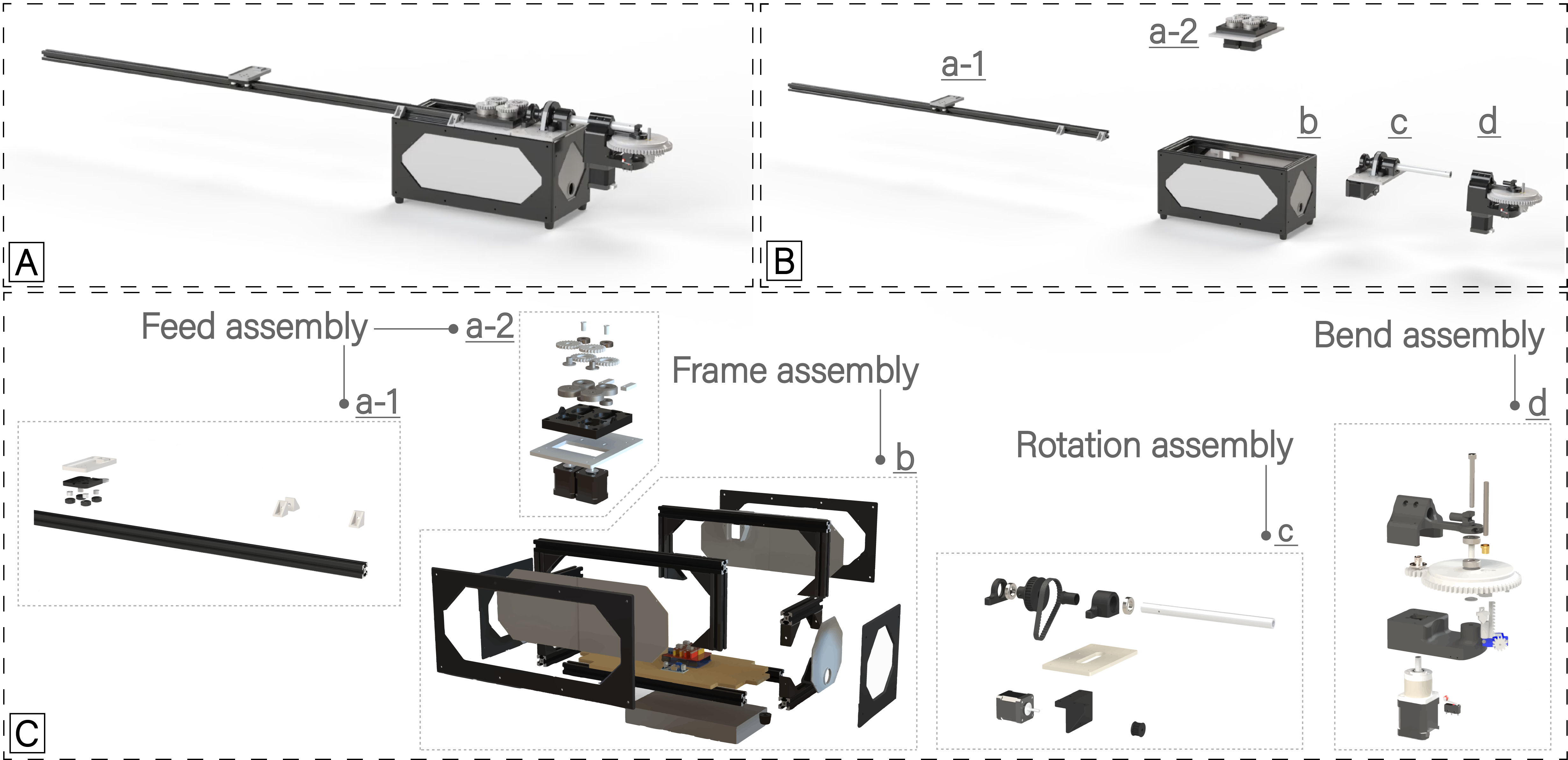}
  \caption{Machine Overview. (A) Full machine assembly. (B) Partial exploded view of the four sub-assemblies. (C) Full exploded view, grouping sub-assemblies. The Feed assembly (a) consists of two disjoint sections; the tail rail (a-1) which secures the rear end of the wire on a sliding gantry, and the feeder (a-2) which feeds/translates the wire forward. The Frame assembly (b) holds all sub-assemblies in place and houses electronics. The Rotation assembly (c) rotates the Bending assembly between bending planes. The Bending assembly (d) physically bends the wire in a single plane.}
  \Description{CAD}
  \label{fig:CAD}
\end{figure*}

The machine is modularly designed as four key sub-assemblies (Figure \ref{fig:CAD}). The Bending Assembly is responsible for bending operations, the Rotation Assembly is responsible for rotating the Bending Assembly into a new bending plane, the Feed Assembly is responsible for feeding material into the bending assembly, and the Frame Assembly holds all sub-assemblies together.


\subsection{Feed Assembly}
\label{Feed Assembly}
The feed assembly is responsible for feeding\textemdash or translating\textemdash the wire forward into the nozzle in precise increments. Feeding is accomplished by two pairs of counter-rotating wheels that contact the wire at a point on their circumferences.

The feeding mechanism consists of 4 wheels driven by two Nema 17 stepper motors. Each motor drives a pair of two wheels, a driving wheel and a driven wheel, placed on either side of the wire. The driving wheel is mounted to the stepper motor via a flange coupling, and drives the driven wheel via 25-toothed gears mounted to each wheel. This arrangement results in the pair of wheels counter-rotating at equal speeds, allowing them to feed wire between them. These wheels are also pre-loaded in compression to grip the wire firmly. Users can adjust this contact force via a sliding mechanism through an M4 screw and nut fixed to the compression mechanism on each pair's driven wheel. The wheels are machined from steel to increase wear resistance and longevity for high contact forces with the wire, and are machined on a lathe and 2-axis mill. 

Finally, the tail rail is a passive mechanism that prevents axial rotation of the wire during feed and bend operations. It accomplishes this by clamping the rear end of the wire, and moves passively with the wire during feed operations. It is comprised of a 4 foot section of 2020 aluminum extrusion (resembling a tail), a V-wheel gantry cart which rolls along this tail, and a 3D printed attachment that allows users to secure the distal end of the wire to the cart. 

\subsection{Rotation Assembly}
\label{RotationAssembly}
The rotation assembly is designed to rotate the bending plane of the bending mechanism. It is these rotations that allow out-of-plane bending operations that permit the creation of 3D structures. 

The rotation assembly is mounted to the frame at one end of a rotating shaft; the bending assembly is mounted to the other end. By rotating this shaft, the rotation assembly rotates the bending plane. The rotation shaft itself is a 16mm OD x 10mm ID aluminum tube of length 220mm, and wire is fed through its center-hole. 

The rotation assembly rotates the rotation shaft along its longitudinal axis using a belt drive. This is achieved using a Nema 17 stepper motor that drives a shaft-mounted hub through a belted 3D-printed gear reduction. A spacer spaces this driven mounting hub from a rear bearing block which allows the shaft to rotate relative the machine frame, while constraining movement in translation. A second, front bearing block is used to counter bending moments from the weight of the bending assembly. 


\subsection{Bending Assembly}
\label{Bender}
The bending assembly constitutes the mechanism responsible for physically bending the wire. It can bend the wire up to 155\degree{}  bi-directionally. The bending assembly is fixed to the rotation shaft via the shaft mount allowing the bending assembly to be rotated. A stepper motor (Nema 17) with a 26.85:1 gearbox is fixed to the shaft mount and drives a 15-tooth gear. This gear drives a large 60-tooth bender gear, with gear ratios chosen to allow bending the wire with 37.9Nm of applied torque. 

The bender gear houses the bending peg, which is the part responsible for physical deformation of the wire. This bending peg is mounted 20.4mm from the center of the bender gear to provide the lever arm necessary for bending. It consists of a steel rod 6.35mm in diameter and 66.4mm in length that physically contacts the wire during bending. This rod can be retracted to allow it to pass under the wire when switching bending direction. The retraction is enabled by a rack-and-pinion mechanism which translates rotary motion from a servo (MG90S) into linear retraction using a rack mounted to the steel rod. To attenuate the stress experienced by the printed gear during bending, the steel bending rod travels through a brass bushing in the bender gear. This bushing reinforces the gear by distributing forces during bending and guides the rod during retraction. The steel bending rod and brass bushing were machined on a lathe and a 2-axis mill.

The bending mechanism also contains a homing sensor that calibrates the bending angle at the start of fabrication. This sensor is divided into two parts, a stationary limit switch attached to the shaft mount, and a trigger mounted under the rotating bending gear which depresses the limit switch during the homing operation. 

Finally, the nozzle itself guides the wire during feeding and bending operations. It is a monolithic part of the shaft mount, and inserts into the shaft during mounting. The inside of the nozzle is tapered to guide the wire through a 3.2mm diameter exit hole. It also contains two steel rods (3mm diameter, 15mm length) that the wire is bent around. These rods set the bending radius of the wire's neutral axis to 3mm and transmit the bending forces more uniformly to the 3D-printed nozzle.




\subsection{Frame Assembly}
\label{Frame Assembly}

The frame assembly is the chassis for the machine, holding all sub-assemblies together and housing the electronics box. The frame is assembled from 12 lengths of 2020 aluminum extrusion and connected using 90-degree angle brackets. Panel covers protect the internal electronics from environmental workshop hazards. The brackets and panels are 3D-printed from PLA.


\subsubsection{Electronics box}

The electronics system consists of an Arduino Uno connected to a CNC shield and 4 DRV8825 stepper motor drivers. These drive the four stepper motors and a servo distributed among the bending, feed and rotation assemblies. The Arduino also senses inputs from the limit switch. The system includes a dedicated fan and power supply, operating at 24V, and the key electronics are housed in a 3D-printed enclosure in the frame assembly.  


The Arduino Uno is the main controller for the wirebender. It is responsible for sending signals to the stepper motor drivers and the servo. The CNC shield is an expansion board for the Arduino Uno that allows for easy connection of stepper motor drivers and other components. It provides the necessary connections for four DRV8825 stepper motor drivers, taken from the Arduino CNC shield V3.0 Kit, which drive the stepper motors with up to 1A of current. These also contain built-in thermal protection that prevents burn-out in the case of inadequate cooling. A 24V-rated fan is placed on top of the 3D printed electronics enclosure to cool the CNC driver board. We use a DC-DC buck converter (LM2596) to step down the 24V input voltage to 5V for the bending servo, which can be driven with up to 3A of current. The power supply for the system is a 24V power supply capable of providing 5A of continuous current to power all system components.

\begin{figure*}[h]
  \centering
  \includegraphics[width=0.99\linewidth]
{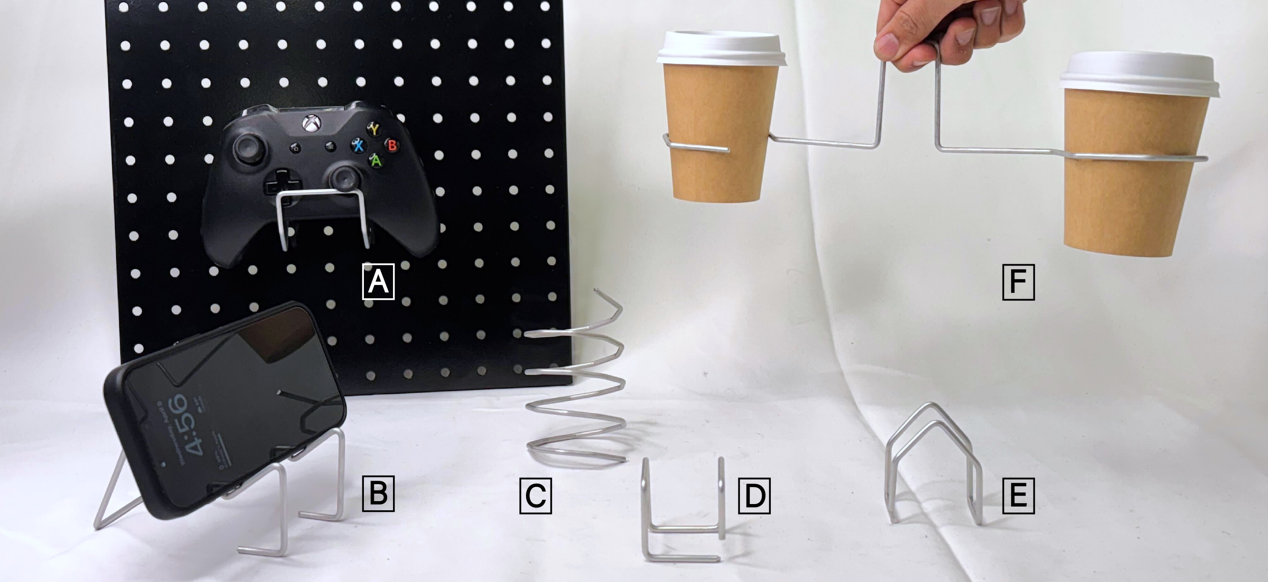}
  \caption{Applications examples designed and fabricated using our system: a (A) pegboard holder for an X-box controller, (B) phone stand, (C) spring, (D) simple cube, (E), house model, and (F) a cup holder.}
  \Description{Applications}
  \label{fig:applications}
\end{figure*}

\subsection{Machine performance}
\label{Assembly}

Having detailed each sub-assembly, we report the key resolution and speed characteristics of our system that arise from its design.

Our machine is designed to feed, bend and rotate with high resolution. We use inexpensive Nema 17 stepper motors across all mechanisms, and increase their native resolution of 1.8\degree{} through microstepping and transmissions. The Feed assembly motor leverages 32-step microstepping to give a feed resolution of 0.0184mm via the feeder wheel diameter of 37.3mm. The Bending assembly motor uses a transmission instead of microstepping to give an angular resolution of 0.017\degree{}. Finally, the Rotation assembly uses 32-step microstepping and a transmission to give a resolution of 0.022\degree{}. As a result of these resolutions, our nominal operating speeds are an angular rate of 15.1\degree/s for bending, an angular rate of 131.3\degree/s for rotation, and a feedrate of 110.2mm/s during feeding. 

All together, our machine is inexpensive, maximizes the use of 3D-printed and COTS components, and is modularly designed into 4 sub-assemblies to permit re-design and further customization.



\section{Applications}

We used our system to design and fabricate a variety of wireframe objects, a sample of which are shown in Figure \ref{fig:applications}.

The cube (Figure \ref{fig:applications}D) has a side length of 30mm and was fabricated as a characteristic example of \system{}'s ability to fabricated objects precisely. It was fabricated in 2 minutes 21 seconds, requiring 241mm of material for a total material cost of \$0.48. The house model (Figure \ref{fig:applications}E) was chosen as a wireframe analogue of a typical aesthetic model that can be found on Thingiverse for 3D printing. It was fabricated in 2 minutes 50 seconds using 292mm of material, costing \$0.58. The cupholder (Figure \ref{fig:applications}F) is designed to carry 2 coffee cups, freeing up a person's 2nd hand for other tasks. It was chosen to illustrate our system's ability to fabricate commonly-used commercial products, and to demonstrate larger scale designs our system can fabricate. It was designed by uploading models of 2x8oz cups to our UI and tracing a wireframe to hold the cups and provide a handle. It was fabricated in 4 minutes 18 seconds using 469mm of material, costing \$0.93. The pegboard mount for the Xbox controller (Figure \ref{fig:applications}A) was chosen to illustrate our ability to create another popular application of aluminum wireframes; creating custom pegboard mounts. It was fabricated in 4 minutes 31 seconds using 380mm of material, costing \$0.76. The phoneholder (Figure \ref{fig:applications}B) was chosen to demonstrate creating wireframe equivalents of other common household objects. It was fabricated in 3 minutes 44 seconds using 362mm of material, costing \$0.72. Finally, the spring (Figure \ref{fig:applications}C) illustrates our machine's ability to approximate curvature in 3D space, and demonstrates our ability to create structures exhibiting compliance. It was fabricated in 9 minutes 8 seconds using 641mm of material, costing \$1.28. 

Combined, these applications illustrate the versatility of our system to rapidly and inexpensively create a variety objects. Any objects fabricated during development of this work that were not kept were recycled through our institution's Facilities department. Having described the breadth of application examples that \system{} is capable of producing, we next evaluate our system.






\section{Evaluation}

In our evaluation, we focus on what we see as key determinants of a successful wirebending machine: the precision with which it can create parts, and the materials with which it can successfully create them. 

\subsection{Torque and wire diameter capabilities}

Our machine must deform wire plastically in order to perform a successful bend operation. In order to determine the torque requirements of our machine, to evaluate the range of material stiffnesses and wire diameters our machine can bend, and to support replication of our study, we conducted tensile tests on our 3mm diameter Aluminum 6061-T6 wire using an Instron 5585 load frame. 

We followed the ASTM E8 standard to execute tensile tests on the same circular specimens used during wirebending. In order to test circular cross-section specimens on the Instron, we threaded the ends of our specimens and connect these to square blocks that can be held in tensile grips. Specimens were machined with a lathe using a single-point turning tool to produce a gauge diameter of 2mm with a gauge length of 8mm in alignment with ASTM standards. During testing, specimens were tensioned at a crosshead speed of 1.27 meters/minute until failure. We conducted tensile tests for N=9 samples and report key mean values. Using the 0.2\% offset method, we find a yield stress of 268.47 MPa at a yield strain of 0.00407, an ultimate tensile strength of 362.14 MPa at a strain of 0.0306, an Elastic Modulus of 68.03GPa and a strain at fracture of 0.0541.

We use the bending equation with a plastic section modulus for the wire's 3mm circular cross-section and the ultimate tensile stress from our Instron results to predict the torque required to bend the wire using our bending peg mounted at a radius of 20.4mm. We found an output torque requirement of 1.63Nm. Using our COTS Nema 17 stepper motors with torque ratings of 0.5Nm and angular resolution of 1.8 degrees, we designed a transmission consisting of a 26.85:1 gearbox and 4:1 external gear. This yielded the angular resolution of 0.017\degree{} previously reported, while raising our output torque to 37.9Nm after accounting for frictional losses and torque loss due to a fall in the moment arm during bending as a result of setback. We verified that this allowed us to execute bends at 15\degree{}/s with a 23x torque margin to permit bending larger diameter or stiffer wire materials in the future. 

We then used 32-step microstepping in conjunction with our transmission to drop the output torque to 1.86Nm, yielding a narrow 14\% margin over the predicted output torque requirement of 1.63Nm. Bending the wire at this setting led to highly unreliable performance, successfully bending individual wire stock, and failing to bend others, revealing that our applied torque was correctly related through our calculations to the tensile strength derived from our instron results. Extrapolating for stresses required to bend stiffer or larger diameter wires, our existing architecture has sufficient torque to bend 3mm steel with a 6x margin, or the same 6061 T6 wire up to 6.8mm in diameter with a 2x torque margin. However, our 3D-printed structure may require strengthening before this torque can be applied.


\subsection{Fabrication accuracy}

\begin{figure}[h]
  \centering
  \includegraphics[width=0.99\linewidth]
{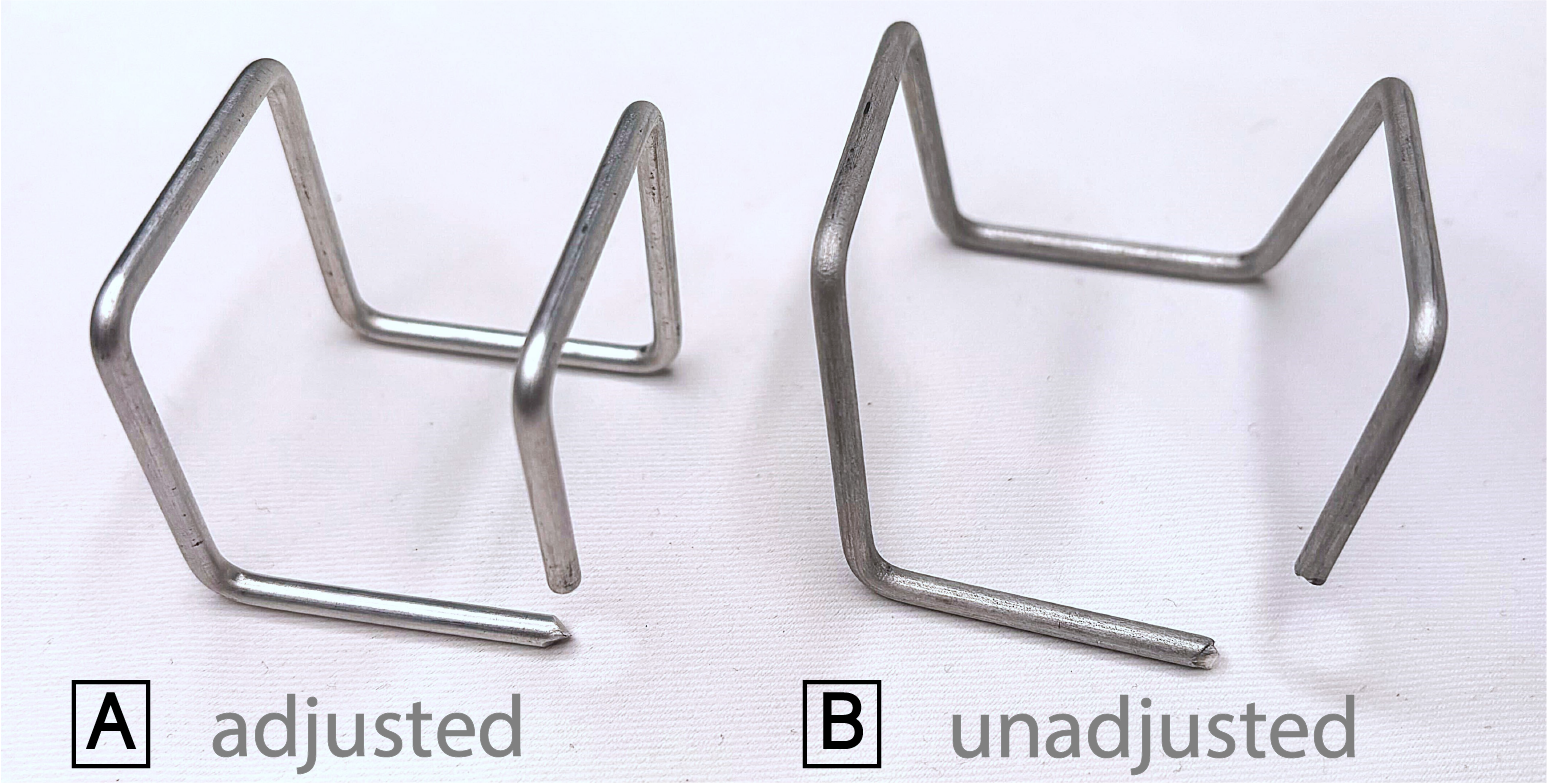}
  \caption{Fabrication accuracy. (A) A cube fabricated using our design tool and error-correction pipeline. (B) The same cube fabricated without error-correction.}
  \Description{Cumulative error}
  \label{fig:cumulative-error}
\end{figure}

To assess our pipeline's ability to accurately fabricate wireframe objects, we evaluated the dimensional accuracy of each of our fabrication procedures: feed, bend, and rotate. A demonstration of our system's ability to error-correct is shown in Figure \ref{fig:cumulative-error}.

\subsubsection{Feed accuracy}
The feeding accuracy dictates how precisely our fabricated wireframes' edge lengths mirror their digital counterparts. We fabricated N=20 samples of U-shaped wireframes, consisting 3 sides of length 35mm with bend angles of 90\degree{}. We fabricated these using both our feed-adjusted commands and commands without adjustment, comparing the results. Without the feed adjustment, we found a mean error of 2.25mm with Std of 0.20mm. After adjusting feed lengths for the same path, we found a mean error of 0.13mm with Std of 0.12mm. This represented a decrease in feed error of 17x using our model, permitting us to create true-to-size wireframes with significantly higher fidelity than without our model. 

\subsubsection{Bending accuracy}

The bending accuracy dictates how precisely our fabricated wireframes' internal angles mirror their digital counterparts. To test this, we first evaluated the fidelity of our individual setback and springback models. We first commanded angles from 0\degree{} to 150\degree{} in 10\degree{} increments using our setback model, using N=5 samples per increment. These setback-compensated angles showed a mean error of 10.23\degree{} with Std of 1.63\degree{} to our commanded angles, for commanded angles greater than 10.23\degree{}. In addition, the error was equal to commanded angles below 10.23\degree{}. This validated both our setback and springback models, as our non-linear setback compensation led to a constant angle error beyond the springback angle of 10.23\degree{}, and a linear relationship below 10.23\degree{}.

\begin{figure}[t]
  \centering
  \includegraphics[width=0.99\linewidth]
{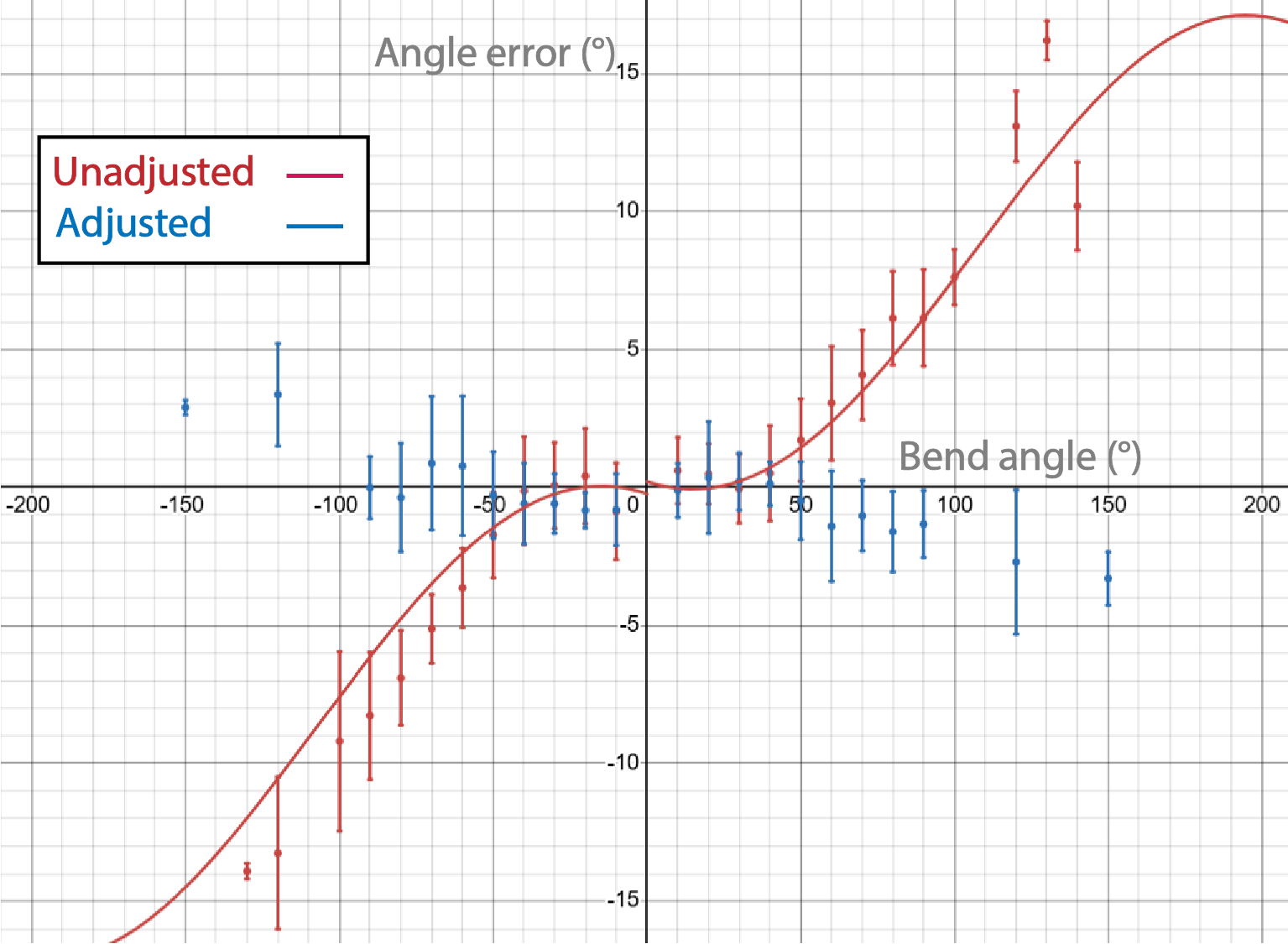}
  \caption{Measurements of bend angle error for different commanded bend angles, comparing error-adjusted vs unadjusted results. Highly non-linear errors affect bend angles for objects fabricated without error-compensation.}
  \Description{Error vs angle}
  \label{fig:ErrorGraph}
\end{figure}

To evaluate the efficacy of our combined model, we commanded bend angles in increments of 10\degree{} using our adjusted model (springback and setback-compensated) and compare this against unadjusted commands, showing the bend angle errors in Figure \ref{fig:ErrorGraph}. The mean error of our adjusted angles was 0.8\degree{} with Std 1.40\degree{}, compared to a mean of 5.34\degree{} with Std 1.67\degree{} for unadjusted values. Our adjusted model on average leads to a 7x decrease in angle error. This is particularly noticeable for large bend angles, where unadjusted bend commands lead to a bending error of 15\degree{}, significantly comprising wireframe accuracy without our model. After calibration, the adjusted mean error did not reach 0, which we believe is primarily due to variability in material microstructure due to the tempering process used in aluminum 6061 T6. 

\subsubsection{Rotation accuracy}
The rotation accuracy dictates how precisely the angles between different planes in our fabricated wireframes mirror their digital counterparts. The rotation accuracy is not affected by the wire; it is instead dictated by the stepper motor resolution and mechanical slack in the rotation assembly. We found a mean error in our rotation angles of 0.05\degree{} with Std 0.88\degree{} for N=20 samples. Combined, our machine architecture and error-correction pipeline affords reliable creation of accurate wireframe structures.

\section{Discussion}


We have introduced a computational design tool and a wirebending machine for fabricating custom wireframe objects. Below, we outline limitations and areas for future work.



\subsection{Wirebending materials} 

We fabricated our objects using Aluminum 6061-T6, one of the most popular forms of aluminum available due to its low cost and ease of recycling. However, approximately 5\% of the samples suffered from brittleness, typically fracturing before bending to 90\degree. This may be attributed to the tempering process used to produce 6061 T6 aluminum which is achieved by precipitation hardening and can lead to impurities in the material causing brittleness. A promising alternative aluminum material may be 6061-O, an annealed variation that may show superior performance to 6061-T6 due to more uniform microstructure. We wirebent 30 specimens of 6061-O aluminum without issue, suggesting this to be a promising material for future exploration. A key motivation for using aluminum in personal wirebending is its recyclability, and while all our stock was recycled externally, future work will explore straightening and spooling previously fabricated wireframes for re-fabrication in-house. 

\subsection{Complexity of fabricated objects} 

Limits on our machine's ability to feed, bend and rotate to any value impede its ability to fabricate arbitrarily complex objects. Our machine can currently execute bends bi-directionally on the interval $\pm[155\degree]$. While there is no minimum absolute bend angle due to our springback compensation, the maximum angle of 155\degree{} is attributed to two compounding constraints: our machine architecture supports bending up to 165\degree{} before wire contacts the rotation shaft, and 10.23\degree{} of springback reduces this to a rounded maximum of 155\degree{}. Future work will explore materials that exhibit less springback, and adjusting the machine design to accommodate larger bends to approach the maximum theoretical bend angle of 180\degree{} where the wire contacts itself. This would significantly impact our ability to fabricate wireframes that do not exhibit "gaps" in their continuity. Another contributor to gaps is our non-zero bending error, which we attribute to material irregularities and possible skipped steps of our stepper motors under high loads. Future work will explore compensating for this using feedback control of bend angles. Finally, we also scripted manual commands to verify our machine's ability to create circular segments as in prior work~\cite{wu2024tune}, and future work will encode this ability programmatically in software.

Our machine can currently execute rotations on the interval $\pm[360\degree]$ before our cabling becomes intertwined. This is not an issue, as geometrically, any angle can be fabricated using a bi-directional rotation interval of just $\pm[90\degree]$ given a bi-directional bending interval of $\pm[180\degree]$. However, ensuring collision-free bending between the wire and bending assembly for more complex shapes may require continuous rotations, and future work will explore slip rings to allow continuous rotations. 

Our machine architecture can currently execute feeds on the interval $[25mm, L-400mm]$, where $L$ is the length of aluminum stock. The maximum feed is reduced by 400mm due to the distance between the feeder wheels and the nozzle, and future redesigns will move these closer to minimize this. The minimum feed is set by the radius swept by the bending peg, and future re-designs will decrease this to allow fabricating higher fidelity wireframes. Our current torque margin of 23x can bend this shorter lever arm without upgrading our bending assembly.

\subsection{Complexity of designed wireframes} 

Path planning for fabricating 3D wireframe structures is a complex problem at the intersection of graph theory, geometry, and physical simulation. The task can be abstracted as finding a path through a graph \( G = (V, E) \) that traverses each edge once to enable fabrication from one continuous length of wire. This is equivalent to the \emph{Chinese Postman Problem}. However, fabricating objects physically requires accounting for physical and geometric constraints introduced by machine limitations and by self-intersections with the wire and machine. This problem is a high-dimensional, nonlinear optimization task in the form of a constrained motion planning problem in 3D space, known to be NP-hard. Given that 3D models are commonly represented as triangulated meshes\textemdash wireframes\textemdash it is possible to search simple models directly for viable wirebending paths. However, the number of paths is exponential in the degree of the wireframe, and future work may address how to plan, represent, and recommend paths to users. In our work, we give control to users to author their own wireframes directly, and build an interface that supports stenciling designs over 3D models, while communicating possible fabrication issues. In our study, we found this presented a useful CAD analog to designing 3D wireframes, and future work will seek to test this empirically through user studies. We will particularly explore and improve collision-checking in the animation, for which we will visualize both wire-wire collisions and wire-machine collisions. Attempting to script low-level machine commands to realize 3D shapes is a challenging task, which our end-to-end pipeline has sought to address. Our full pipeline for designing 3D wireframe designs, checking fabricability, addressing errors, and automatically fabricating them is documented to support more work in this area.

\section{Conclusion}

In this paper, we introduced a desktop wirebending machine and computational design tool for creating 3D wireframe structures. We have illustrated that they allow users to rapidly and inexpensively create custom 3D wireframe structures from 3mm aluminum wire. The design tool allows users to generate wireframe designs and assess their fabricability, before a path-planning procedure converts designs into fabrication instructions. Our wirebending machine costs \$293 in parts and can form aluminum wire into 3D wireframe structures through an ordered sequence of feed, bend, and rotate instructions. Our technical evaluation reveals our system's ability to account for material elasticity and kinematic error sources to produce accurate 3D structures from inexpensive hardware. Our software is written in freely available software and is documented together with complete documentation of our machine. 

\bibliographystyle{ACM-Reference-Format}
\bibliography{references}

\appendix









\end{document}